\documentclass[
  reprint,
  superscriptaddress,
  amsmath,amssymb,
  aps,
  pra,
]{revtex4-2}%\documentclass[twocolumn,prb,showpacs,multicol,amsmath,amssymb]{revtex4-1}
\usepackage{graphicx}
\usepackage[caption=false]{subfig}
\usepackage{blindtext}
\usepackage{lipsum}
\usepackage{sidecap}
\usepackage{epstopdf}
\usepackage{gensymb}
\usepackage{soul}
\usepackage{color}
\usepackage{amssymb}
\usepackage{amsmath}
\usepackage{bm}% bold math
\graphicspath{{Figures/}}
\usepackage{times}
\usepackage{physics}
\usepackage{enumitem}
\usepackage[normalem]{ulem}
\usepackage{soul}
\usepackage{sidecap}

\newcommand{\be}{\begin{equation}}
\newcommand{\ee}{\end{equation}}
\newcommand{\bk}{{{\bf{k}}}}
\newcommand{\bq}{{{\bf{q}}}}

\newcommand{\bea}{\begin{eqnarray}}
\newcommand{\eea}{\end{eqnarray}}

\newcommand{\bd}{\begin{displaymath}}
\newcommand{\ed}{\end{displaymath}}
\newcommand{\ba}{\begin{array}}
	\newcommand{\ea}{\end{array}}
\newcommand{\bi}{\begin{itemize}}
	\newcommand{\ei}{\end{itemize}}
\newcommand{\bc}{\begin{center}}
	\newcommand{\ec}{\end{center}}
\newcommand{\bfl}{\begin{flushleft}}
	\newcommand{\efl}{\end{flushleft}}
\newcommand{\bfr}{\begin{flushright}}
	\newcommand{\efr}{\end{flushright}}
\newcommand{\no}{\nonumber}

\newcommand{\bl}{\begin{aligned}}
	\newcommand{\el}{\end{aligned}}
\def\bk{{\bf k}} \def\bq{{\bf q}}  
  \def\bd{{\bf d}}

\def\6{\partial}

\def\o{\omega}

\def\={\!\!\!&=&\!\!\!}
\def\+{\!\!\!&&\!\!\!+~}
\def\-{\!\!\!&&\!\!\!-~}

\newcommand\redout{\bgroup\markoverwith{\textcolor{red}{\rule[.5ex]{2pt}{0.4pt}}}\ULon}

\usepackage[colorlinks=true,citecolor=blue]{hyperref}
\hypersetup{colorlinks=true,citecolor=blue,linkcolor=red,urlcolor=blue}

\graphicspath{{.}{./new_figs/}}

\begin{document}
\title{
%Role of Fermi Surface Geometry and van Hove singularity on Optical Response of Superconductor Sr$^{}_2$RuO$^{}_4$}
Influence of Fermi Surface Geometry and Van Hove Singularities on the Optical Response of Sr$_2$RuO$_4$
}
\author{Meghdad Yazdani-Hamid}
%\email{meghdad.yazdani@abru.ac.ir}
\affiliation{Department of Physics, Bu-Ali Sina University, 65178, 016016, Hamedan, Iran}
%\affiliation{Department of Physics, Ayatollah Boroujerdi University, Boroujerd, Lorestan 65151-36111, Iran}
%
\author{Mehdi Biderang}
%\email{???}
\affiliation{Department of Physics, University of Toronto, 60 St. George Street, Toronto, Ontario, M5S 1A7, Canada}
\affiliation{DelQuanTech Inc., 500 Doris Ave., Toronto, Ontario, M2N 0C1, Canada}
\author{Alireza Akbari}
\affiliation{Beijing Institute of Mathematical Sciences and Applications (BIMSA), Huairou District, Beijing 101408, China}
\date{\today}
%
%%%%%%%%%%%%%%%%%%%%%%%%%%%%%%%%%%%%%%%%%%%%%%%%%%%%%%%%%%%%%%%%%%%%%%%%%%%
\begin{abstract}
{

Motivated by the sensitivity of Sr$_2$RuO$_4$ to Fermi surface reconstructions under strain,
we investigate how Fermi surface geometry and Van Hove singularities influence the optical Hall
response and polar Kerr effect. Within a three-orbital model, we explore the impact of chemical
potential and interlayer hopping on superconducting pairing and response functions. We find that
$d_{x^2-y^2}$ and $d_{x^2-y^2}+ig$ symmetries are the leading candidates for the quasi-2D orbital,
while a chiral $p$-wave state in the quasi-1D orbitals is essential for generating an accessible Kerr
angle. The Lifshitz transition is shown to affect coherence factors and density-of-states peaks,
producing sharp signatures in $T_c$ and optical transport. Inter-orbital charge transfer further
enhances these effects by modifying the balance between quasi-1D and quasi-2D contributions.
These results provide a framework for interpreting Kerr effect experiments in
multi-orbital superconductors.
}

\end{abstract}
%%%%%%%%%%%%%%%%%%%%%%%%%%%%%%%%%%%%%%%%%%%%%%%%%%%%%%%%%%%%%%%%%%%%%%%%%%%
%
\maketitle
%

%
%%%%%%%%%%%%%%%%%%%%%%%%%%%%%%%%%%%%%%%%%%%%%%%%%%%%%%%%%%%%%%%%%%%%%%%%%%%
%%%%%%%%%%%%%%%%%%%%%%%%%%%%%%%%%%%%%%%%%%%%%%%%%%%%%%%%%%%%%%%%%%%%%%%%%%%
%%%%%%%%%%%%%%%%%%%%%%%%%%%%%%%%%%%%%%%%%%%%%%%%%%%%%%%%%%%%%%%%%%%%%%%%%%%

\section{Introduction}

Strontium ruthenate, a transition metal oxide with enigmatic superconducting symmetry,
continues to captivate researchers due to its intricate interplay of orbital physics,
electronic structure, and symmetry-breaking phenomena~\cite{maeno1994superconductivity,Mackenzie2003superconductivity}.
As a prime candidate for studying strongly correlated electron systems, Sr$_2$RuO$_4$
provides a unique platform to explore the mechanisms behind superconductivity and its
potential applications in advanced technologies~\cite{Barber2019Piezoelectric,jerzembeck2022superconductivity,Hicks2024Probing}.
Achieving a detailed understanding of its electronic behavior remains a central goal,
with particular emphasis on how variations in Fermi surface topology, chemical potential,
and interorbital coupling shape its superconducting and transport properties~\cite{Mackenzie2017,Maeno2024}.
External perturbations such as strain can significantly tune the electronic structure,
modifying the density of states (DOS), inducing Lifshitz transitions, and enhancing the
superconducting transition temperature~\cite{Hicks2014Strong,Steppke2017Strong,Jerzembeck2024Tc}.
They can also influence the magnetic properties by adjusting the Ru–O–Ru bond angles
and distances, thereby impacting exchange interactions and the spin-fluctuation
spectrum~\cite{Cobo2016Anisotropic,pustogow2019constraints,Luo2019Normal,Romer2020Theory,Chronister2022Tuning,Jerzembeck2023Upper}.
These effects, driven by alterations in the DOS and Fermi surface topology, underscore
the potential of external perturbations to uncover and manipulate the fundamental physics
of this material~\cite{Hicks2014Piezoelectric}.
Modifications of the Fermi surface topology can also be achieved by rotating the RuO$_6$
octahedra~\cite{Matzdorf2000Ferromagnetism,Veenstra2013Determining}, increasing the chemical
potential~\cite{IMAI2015Effect}, substituting La for Sr~\cite{Shen2007Evolution,Nishinakayama2024Negatively},
or constructing SrRuO$_3$–SrTiO$_3$ heterostructures~\cite{Kim2020heterostructure}.

A variety of experimental observations suggest the presence of time-reversal symmetry
breaking (TRSB) in the superconducting state, pointing to the condensation of multiple
superconducting order parameters, likely associated with two-dimensional irreducible
representations of the D$_{4h}$ crystal symmetry group~\cite{Mackenzie2003superconductivity,Mackenzie2017,Kallin_2012,Maeno2012Evaluation}.
One of the key pieces of evidence for TRSB pairing is the observation of Hall-type effects,
which involve the generation of a transverse voltage in response to an applied current,
even in the absence of an external magnetic field.
Closely related is the polar Kerr effect, in which circularly polarized light incident
on the superconducting sample is reflected with its polarization rotated~\cite{Xia2006High}.
Despite these findings, the precise origins of both the Hall-type and Kerr effects remain
a topic of ongoing debate, closely tied to questions about the dominant superconducting
orbitals.
Extensive theoretical studies have investigated Hall transport and the Kerr effect using
diverse approaches~\cite{Kim2008Hall,Goryo2008Impurity,Lutchyn2009Frequency,Taylor2012Intrinsic,
Wysoki2012Intrinsic,taylor2013anomalous,Gradhand2013Kerr,Robbins2017Effect,Konig2017Kerr,
Denys2021Origin,Liu2023Impact,Yazdani2024Polar,Zhang2024Quantum}, motivated by the
experimental evidence~\cite{Xia2006High}.
It has been shown that the Hall response originates primarily from TRSB in the gap
functions of the quasi-1D orbitals~\cite{Yazdani2024Polar}, with SOC playing a key role
in orbital mixing, while the quasi-2D orbital contributes only weakly.
Based on these findings, the present work examines how the Lifshitz transition,
associated with the quasi-2D orbital, influences the Hall response that is governed by
the quasi-1D orbitals.

Here, we aim to uncover the mechanisms behind the superconductivity of strontium ruthenate
and explore ways to tailor its properties~\cite{Ghosh2022Strong,Noad2023Giant,Yang2023Probing,Abarca2023Hierarchy}.
Our primary focus is on how variations in the Fermi surface of the $\gamma$ band,
especially its proximity to the Brillouin zone boundary, influence the coherence factors,
density of states, and Hall transport.
We employ a two-dimensional three-orbital tight-binding model combined with a self-consistent
Bogoliubov–de Gennes approach. These effects are studied by tuning the chemical potential
and enhancing the coupling between the quasi-2D and quasi-1D orbitals. This coupling is
modeled through $z$-direction hopping, which at specific values produces nearly degenerate
states near the Fermi level, thereby altering the fillings of the corresponding bands.
We begin by examining the leading pairing symmetries of the $d_{xy}$ orbital within the
weak-coupling regime.
 Based on our calculations of the dependence of $T_c$ on the
$z$-direction hopping and chemical potential, we find that the favored pairing symmetries
in the quasi-2D orbital are $d_{x^2-y^2}$ and $d_{x^2-y^2}+i g_{xy(x^2-y^2)}$.

Next, we turn to the core theme of this paper: the impact of the Lifshitz transition on Hall
transport driven by the quasi-1D orbitals.
In the following section, we introduce the physical model used to describe the electronic
band structure and superconducting ground state, highlighting the favored pairing channel
on the quasi-2D orbital within the weak-coupling limit.
Section~\ref{OPH} focuses on calculating the dynamical Hall conductivity using the
superconducting gap functions derived in Sec.~\ref{Sec:Model}.  We show that the optical
responses of these leading pairing symmetries are very similar because they share identical
gap nodes, indicating that nearest-neighbor pairing dominates in the $d_{xy}$ orbital. This
further implies that the gap function of this orbital does not need to be purely complex.
Our results demonstrate that the dominant Hall response originates from quasiparticle spectra
away from the Fermi level, with only negligible contributions from low-energy states close to it.
This behavior persists even at finite temperatures. Moreover, as the $z$-direction hopping
increases, the DOS of the $d_{xy}$ orbital near the Fermi level decreases, reducing the
differentiation between the $\beta$ and $\gamma$ bands until they become nearly degenerate.
This proximity in the diagonal region enables both bands to contribute comparably to Hall
transport, thereby enhancing the Kerr effect. In contrast, spin–orbit coupling suppresses this
$\beta$–$\gamma$ near-degeneracy for all $k_z$ values, resulting in a reduction of the Kerr signal.
Increasing the chemical potential enhances the Kerr angle up to a critical value, where the
$\gamma$ sheet becomes open; this enhancement can be attributed to the amplification of the
DOS peak near the van Hove singularity. In Sec.~\ref{PKE}, we calculate the polar Kerr angle
using a two-orbital model for the longitudinal conductivity, following the multigap framework
of Ref.~\cite{dora2003optical}, which reflects the tetragonal crystal symmetry and the distinct
orbital characters of the three Fermi surfaces. Including electron–electron scattering, we show
that scattering further increases the Kerr angle, though the enhancement remains smaller than
that observed in the high-frequency limit.
Finally, the concluding section summarizes the key findings and closes the paper.
%

%
%%%%%%%%%%%%%%%%%%%%%%%%%%%%%%%%%%%%%%%%%%%%%%%%%%
%%%%%%%%%%%%%%%%%%%%%%%%%%%%%%%%%%%%%%%%%%%%%%%%%%
%
%
\begin{SCtable*}
\begin{tabular}{| l | l|} % Change c to l for left justification
\hline
\quad\quad Pairing  & \quad\quad\quad\quad\quad\quad\quad\quad
Momentum Dependence
\\
%\hline
\hline
(a) $d + ig$  & $\Delta_{xy}^{\bk}=  [\Delta^\prime_{xy}(T) + i \Delta^{\prime\prime}_{xy}(T) \sin k_x \sin k_y]
(\cos k_x - \cos k_y)$ \\
\hline
(b) $d_{x^2-y^2}$  &  $\Delta_{xy}^{\bk}=\Delta^\prime_{xy}(T)(\cos k_x - \cos k_y)$ \\ \hline
(c) $p + ip$ & $\Delta_{xy}^{\bk}=  \Delta^\prime_{xy}(T) \sin k_{x} + i \Delta^{\prime\prime}_{xy}(T) \sin k_{y}$ \\ \hline
(d) $s^\prime + id_{xy}$ & $\Delta_{xy}^{\bk}=  \Delta^\prime_{xy}(T)(\cos k_x + \cos k_y) + i \Delta^{\prime\prime}_{xy}(T) \sin k_x \sin k_y$ \\ \hline
(e) $s^\prime + id_{x^2-y^2}$ & $\Delta_{xy}^{\bk}=  \Delta^\prime_{xy}(T)(\cos k_x + \cos k_y) + i \Delta^{\prime\prime}_{xy}(T)(\cos k_x - \cos k_y)$\; \\ \hline
\end{tabular}
\caption{
	The most favored forms of the  quasi-2D orbital superconducting order parameter, $\Delta_{xy}^{\bk }$, for
	(a) $d + ig$: $d_{x^2-y^2}+ig_{xy(x^2-y^2)}$,
	(b) $d_{x^2-y^2}$,
	(c) $p + ip$: \; $p_x + ip_y$ ,
	(d) $s^\prime + id_{xy}$,
	and
	(e) $s^\prime + id_{x^2-y^2}$.
}\label{tab:order_parameters}
\end{SCtable*}

%%%%%%%%%%%%%%%%%%%%%%%%%%%%%%%%%%%%%%%%%%%%%%%%%%%%%%%%%%%%%%
\section{Model Hamiltonian}
\label{Sec:Model}
The Hamiltonian of a 2D systems spread in $xy$-plane describing the normal state of strontium ruthenate is given by\
${\cal H}_{\rm N}  ={\cal H}_{\rm 0}  +{\cal H}_{\rm SOC}$.
Here,
%%%%
\be
\bl
{\cal H}_{\rm 0}  = \sum_{\bk, \nu, \nu', \sigma} \left( \xi^{\bk}_{\nu} \delta_{\nu,\nu'} + {\cal T}^{\bk}_{\nu, \nu'} \right) d_{\bk, \nu, \sigma}^\dagger d_{\bk, \nu', \sigma}^{}{\color{blue},}
\el
\ee
resulting in a Fermi surface composed of three bands: a hole-like $\alpha$ band from quasi-1D orbitals and two electron-like $\beta$ and $\gamma$ bands, derived from quasi-1D and quasi-2D orbitals, respectively~\cite{Mackenzie2003superconductivity}.
Here
$ \nu, \nu' \in \{xz, yz, xy\}$
index the $ t_{2g} $ orbitals, where
$d_{\bk, \nu, \sigma}^\dagger  \; (d_{\bk, \nu, \sigma} )$ is the fermionic creation  (annihilation) operator for the orbital state $\nu$ with momentum $\bk$ and spin $\sigma$.
Moreover,
$  \xi^{\bk}_{\nu} $ represents the dispersion of  orbital $ \nu $, %and
$ {\cal T}^{\bk}_{\nu \nu'} $ describes the hybridization between orbitals $ \nu $ and $ \nu' $, and
$\delta_{\nu,\nu'}$
is the Kronecker delta function.
 In contrast to the nearly $ k_z $-independent $ d_{xy} $ orbital, the $ d_{xz} $ and $ d_{yz} $ orbitals exhibit significant $ k_z $-dispersion, leading to orbital mixing that affects the $\beta$ and $\gamma$ bands at specific $ k_z $ values~\cite{Haverkort2008Strong,Gingras2019Superconducting}.
The electronic dispersions at this $ k_z $ value, calculated via the tight-binding method, are given by
%%%%%%%%%%
\be
\begin{aligned}
&\xi_{xz}^{\bk} = -\mu - 2t_{xz}^{x} \cos k_x - 2t_{xz}^{y} \cos k_y, \\
&\xi_{yz}^{\bk} =  -\mu - 2t_{yz}^{x} \cos k_x - 2t_{yz}^{y} \cos k_y, \\
&\xi_{xy}^{\bk} =  -\mu - 2t_{xy}^{x} \cos k_x - 2t_{xy}^{y} \cos k_y
%\\&
-
4t_{xy}^{xy} \cos k_x \cos k_y,
\\
&
{\cal T}_{xz,yz}^{\bk} =  -4g \sin k_x \sin k_y
,\quad
{\cal T}^{\bk}_{ xz,xy }
=\!
-8g^\prime \! \cos \frac{k_{x}}{2}
\sin \frac{k_{y}}{2},
\\
&
{\cal T}^{\bk}_{ yz,xy }
=\!
-8g^\prime \! \cos \frac{k_{y}}{2}
\sin \frac{k_{x}}{2},
\end{aligned}
\ee
%%%%%%%%%%
where $ \mu $ is the chemical potential, and $ t^{\vartheta=x,y,xy}_\nu $ are the hopping parameters for orbital $ \nu $. The parameters $ g $ and $ g^\prime $ characterize the intra-quasi-1D and inter-quasi-1D/2D orbital hopping strengths, respectively.
The equilibrium   tight-binding parameters are set to
$
(t^x_{xz}, t^y_{xz}, t^x_{xy}, t^{xy}_{xy}, \mu, g, g^\prime) = (110, 7, 82, 37, 142, 9, 1)\:\text{meV}
%(t^x_{xz}, t^y_{xz}, t^x_{xy}, t^{x,y}_{xy}, \mu, g, g^\prime) = (0.11, 0.007, 0.082, 0.037, 0.142, 0.009, 0.001)\:\text{eV}
$,
and we consider  $  t^y_{yz}=t^x_{xz} $, $ t^x_{yz}=t^y_{xz}  $, and $ t^y_{xy} =t^x_{xy} $~\cite{Gradhand2013Kerr}.

In addition, the atomic spin--orbit coupling (SOC) is included as~\cite{Cobo2016Anisotropic}
\be
\mathcal{H}_{\mathrm{SOC}}
= \sum_{\bk, \sigma}
\mathbf{d}_{\bk, \sigma}^{\dagger} \, H_{\mathrm{SOC}} \, \mathbf{d}_{\bk,\bar{ \sigma}},
\ee
where
$\mathbf{d}_{\bk\sigma}^{\dagger} =
(d_{\bk,yz,\sigma}^{\dagger},
      d_{\bk,xz,\sigma}^{\dagger},
       d_{\bk,xy,\sigma}^{\dagger}
            )$
is the orbital--spin basis vector, and $\bar{\sigma}=-\sigma$.
The SOC operator takes the standard form
$
H_{\mathrm{SOC}} = \lambda \, \mathbf{L} \cdot \mathbf{S},
$
where $\lambda$ is the SOC strength, $\mathbf{L}$ is the orbital angular momentum operator acting in the $\{ d_{yz}, d_{xz}, d_{xy} \}$ basis, and
$\mathbf{S} = \tfrac{1}{2}\boldsymbol{\sigma}$ represents the spin operator with Pauli matrices
$\boldsymbol{\sigma} = ( \sigma_x, \sigma_y, \sigma_z )$.
In matrix form, the SOC contribution in this orbital basis reads
\begin{equation}
H_{\mathrm{SOC}} =
\frac{\lambda}{2}
\begin{pmatrix}
0 & i\sigma_z & -i\sigma_y \\
-i\sigma_z & 0 & i\sigma_x \\
i\sigma_y & -i\sigma_x & 0
\end{pmatrix},
\end{equation}
where each entry is a $2 \times 2$ block in spin space.
Here, we focus on $k_z = \pi$, where the overlap between the $d_{xy}$ and $d_{xz/yz}$ orbitals is maximized and SOC effects are minimal~\cite{Haverkort2008Strong, Oda2019Paramagnetic}. Therefore, we neglect the SOC in the following, although we explicitly note its presence.

\begin{figure}[t]
\centering
\includegraphics[width= \linewidth]{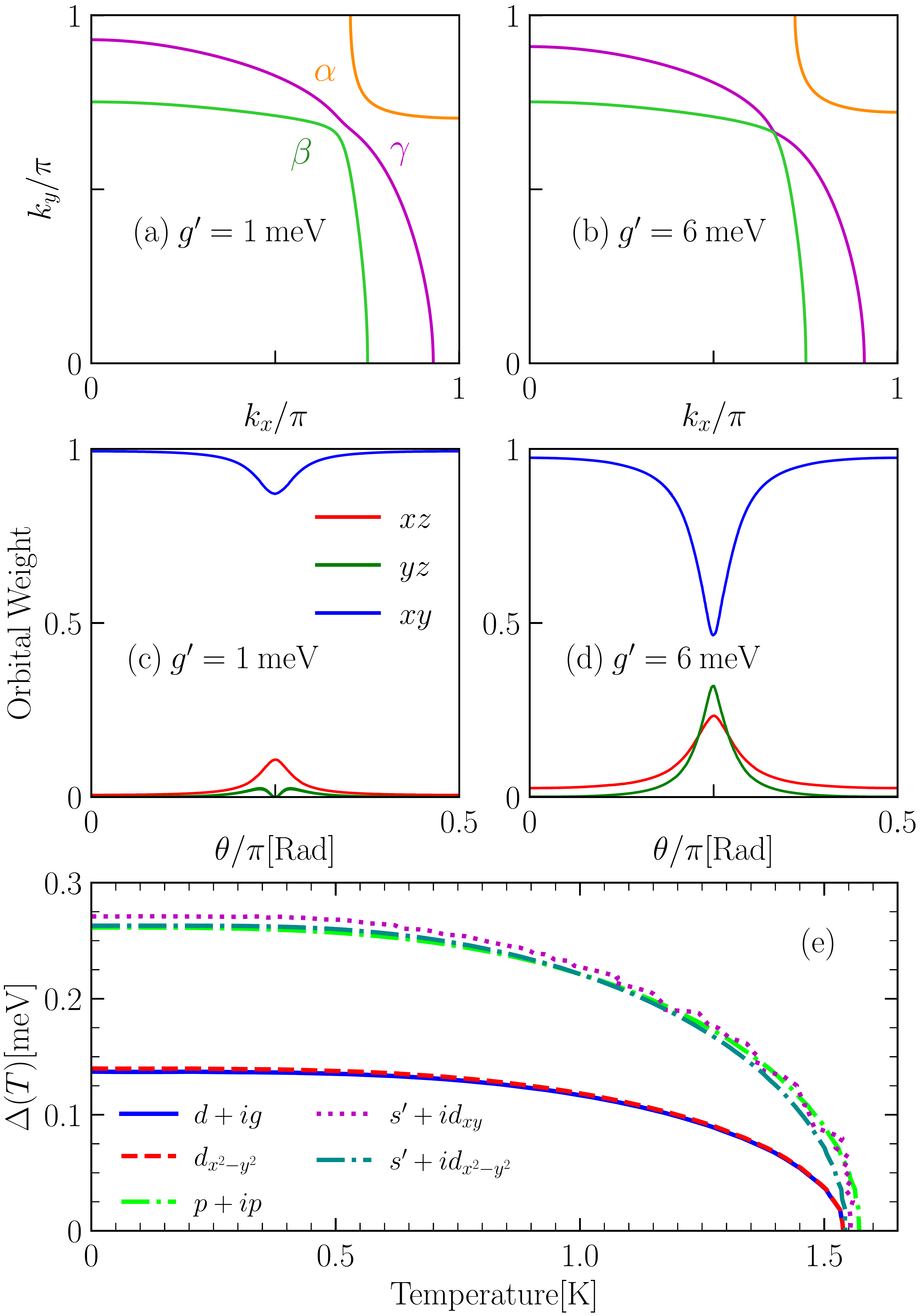}
%\vspace{-1cm}
\caption{
Illustration of the effect of the parameter $g'$ on the Fermi surface: (a) $g' = 1\text{meV}$, and (b) $g' = 6\text{meV}$, highlighting the touching of the two bands, $\beta$ and $\gamma$, in the diagonal region, which predominantly contributes to the Berry curvature and Hall transport. (c) and (d) depict the variations in orbital weight on the $\gamma$-sheet along a quarter of the Brillouin zone for the respective values of $g'$, where $0<\theta=\arctan (k_y/k_x)<\pi/2$. (e) Temperature dependence of the superconducting gap for different order parameters.
}
\label{gapfunc}
\end{figure}

The chemical potential  is crucial for applying biaxial strain. An increase in chemical potential causes  van Hove singularities (vHS) in the [100] and [010] directions, leading to a Lifshitz transition. This also expands the $\gamma$ sheet due to changes in orbital populations.
While increasing the chemical potential indicates positive biaxial strain and the filling of the $\gamma$ band, increasing $g^\prime$ moves the $\gamma$ band away from the vHS near the Fermi level. These parameters influence the proximity of the $\beta$ and $\gamma$ sheets and the orbital weights from $d_{xy}$ and $d_{xz/yz}$ across the Fermi surface. Fig.~\ref{gapfunc}(a) and Fig.~\ref{gapfunc}(b) show that increasing $g^\prime$ significantly alters the diagonal region of Brillouin zone, causing the $\gamma$ and $\beta$ sheets to touch and increasing the separation between them and the hole-like $\alpha$ sheet.
Such a close proximity between $\beta$ and $\gamma$ bands and the emergence of concavity on the  $\gamma$ isotropic band along the diagonal zone, has been observed at $k_z=\pi$ under c-axis uniaxial stress \cite{jerzembeck2022superconductivity}, resulting in decrease of the superconducting transition temperature. In contrast to the case of 2D, such a stress affects the hopping parameters of the $d_{xz/yz}$ orbitals. An increase in $g^\prime$ also brings about the reduction (increase) of the contribution of the $d_{xy}$ quasi-2D orbital ($d_{xz}$ and $d_{yz}$ quasi-1D orbitals) to the $\gamma$ sheet, see plots Fig.~\ref{gapfunc}(c) and Fig.~\ref{gapfunc}(d).
The Hall response is influenced by the one-dimensional nature of the $d_{xz/yz}$ orbitals, making $g^\prime$ important for Hall transport. Specifically, higher $g^\prime$ leads to electron transfer from $d_{xy}$ to $d_{xz/yz}$, decreasing the DOS near the Fermi level.

%%%%%%%%%%%%%%%%%%%%====== figure ======%%%%%%%%%%%%%%%%%%%
%%%%%%%%%%%%%%%%%%%%%%%%%%%%%%%%%%%%%%%%%%%%%%%%%%
%%%%%%%%%%%%%%%%%%%%====== figure ======%%%%%%%%%%

%
The multi-band structure leads to different superconducting gap textures~\cite{Scaffidi2023Degeneracy}, therefore, the superconducting Hamiltonian can be written as
%
%=============================================================
%========================Equation=============================
%=============================================================%
\be
\bl\label{Nammbu}
{\cal H}_{\rm SC} = {\cal H}_{\rm N} +
\sum_{\bk \nu \nu' \sigma \sigma'}
 \left( \Delta_{\nu \nu'}^{\bk} d_{-\bk, \nu \sigma}^\dagger d_{\bk, \nu' \sigma'}^\dagger + \text{h.c.} \right),
\el
\ee
in which the elements of the superconducting order parameter are
obtained through the solution of the self-consistency equation
%
%
%\small
\begin{align}\label{gg}
\Delta_{\nu\nu^\prime}^{\bk}
%=\Delta_{\nu\nu^\prime}({\bk})
= -T \sum_{\omega_n} \sum_{\substack{\nu_{1}, \nu_{1}^{\prime}}} \sum_{\textbf{k}^\prime}
{\rm g}_{\nu\nu^\prime, \nu_{1} \nu_{1}^{\prime}}^{\bk,\bk^\prime}
\,
F_{\nu_{1} \nu_{1}^{\prime}}(\bk^\prime, \omega_n).
\end{align}
%\normalsize
%
%
Here, $\omega_n = (2n + 1) \pi T$ is the fermionic Matsubara frequency, and $F_{\nu \nu'}$ is the anomalous Green's function component.
The superconducting
 coupling matrix
${\rm g}_{\nu\nu^\prime, \nu_{1} \nu_{1}^{\prime}}^{\bk,\bk^\prime}$ defines the pairing interactions,  and describes the orbital part of the  coupling processes in momentum space.
Specifically, this work focuses on the case of
$
{\rm g}_{\nu\nu, \nu_{1} \nu_{1}}^{\bk,\bk^\prime}
=
{\rm g}_{\nu, \nu_{1}   }^{\bk,\bk^\prime}
$,
which indicates the intra-orbital pairing within a diagonal matrix format $\Delta_{\nu\nu'}^{\bk } = \Delta_{\nu}^{\bk }$.
The primary justification for ignoring the consideration of interorbital pairing is its entirely negligible contribution to Hall transport across all frequency regimes.
In this case, we can define the coupling matrix in a separable expression of
\begin{align}
{\rm g}_{\nu, \nu_{1}}^{\bk,\bk^\prime}
=
%\sum_{\substack{\nu=xz, \\ yz,xy}}
\mathtt{g}^\prime_{\nu\nu_1}\textrm{f}^\prime_{\nu}(\mathbf{k})\textrm{f}^\prime_{\nu_1}
(\mathbf{k}^\prime)
+
\mathtt{g}^{\prime\prime}_{\nu\nu_1}\textrm{f}^{\prime\prime}_{\nu}(\mathbf{k})\textrm{f}^{\prime\prime}_{\nu_1}(\mathbf{k}^\prime),
\end{align}
with form factors
$\textrm{f}^{}_{\nu}(\mathbf{k})=\textrm{f}^{\prime}_{\nu}(\mathbf{k})+i \textrm{f}^{\prime\prime}_{\nu}(\mathbf{k})$ for the orbital
$\nu$, encoding the momentum dependence of the interaction~\cite{Nomura2002Detailed,Akbari2013Multiorbital,Zinkl2019Superconducting,Yu2020}.
It represents intra-orbital Cooper pairs, where pairing occurs within the same orbital ($\nu = \nu_1$) or involves hopping to a different orbitals ($\nu \neq \nu_1$)~\cite{Salamone2023High}.

For Sr$_2$RuO$_4$ with its two-dimensional square lattice, the gap symmetry can be classified according to the irreducible representations (irreps) of the D$_{4h}$ point group: extended $s$-wave ($s^\prime$), $d_{x^2-y^2}$, $d_{xy}$, $g_{xy(x^2-y^2)}$, and $p_{x}(p_{y})$. The tetragonal symmetry of the material, together with the strongly one-dimensional character of the $d_{xz/yz}$ orbitals and the two-dimensional nature of the $d_{xy}$ orbital, has motivated numerous proposals for its superconducting state. Some scenarios suggest that the quasi-1D and quasi-2D orbitals share the same pairing symmetry \cite{Raghu2010Hidden,Firmo2013Evidence}, while others advocate different pairings on these orbitals \cite{Sharma2020Momentum}.
 We focus on a time-reversal-symmetry-breaking phase with chiral character on the quasi-1D orbitals, which belongs to the E$_u$ irrep of the D$_{4h}$ group and is consistent with the absence of edge currents as well as with Kerr-effect measurements \cite{Raghu2013Theory,Yazdani2024Polar}.
Thus, the superconducting order parameter can be expressed as $\Delta_{\nu\nu}^{\bk } = \Delta_{\nu}^{\bk } = \Delta^{\prime}_{\nu}(T) \, \textrm{f}^{\prime}_{\nu}(\bk ) +  i  \Delta^{\prime\prime}_{\nu}(T) \, \textrm{f}^{\prime\prime}_{\nu}(\bk )$.
Correspondingly, for $\Delta_{xz}^{\bk}$ and $\Delta_{yz}^{\bk}$, we can consider $(\textrm{f}^{\prime}_{xz}(\bk ), \textrm{f}^{\prime\prime}_{xz}(\bk ), \textrm{f}^{\prime}_{yz}(\bk ), \textrm{f}^{\prime\prime}_{yz}(\bk )) = (\sin k_x, 0, 0, \sin k_y)$ with $\Delta^{\prime}_{xz}(T) = \Delta^{\prime\prime}_{yz}(T)$~\cite{Denys2021Origin,Yazdani2024Polar},
however, for $\Delta_{xy}^{\bk}$, we analyze the preferred forms of the orbital order parameter presented in Table~\ref{tab:order_parameters}~\cite{Hicks2014Strong,Steppke2017Strong,kivelson2020proposal,Romer2019Knight,Furusaki2001Spontaneous}.

Fig.~\ref{gapfunc}(e) shows the BCS prediction for the temperature-dependent order parameter in the $d_{xy}$ orbital, $\Delta^{\prime}_{xy}(T)$, across the considered forms.
The focus here is on the $d_{xy}$ orbital, as variations in the coupling $g^\prime$ and chemical potential impact its DOS near the Fermi level.
We determine the magnitude of $\Delta_0$,  the gap amplitude at zero temperature, by adjusting the pairing strengths in the channels listed in Table~\ref{tab:order_parameters} to yield a transition temperature in line with experimental values, $T_c= 1.5$K~\cite{maeno1994superconductivity}.
The calculated values for the $p+ip$, $s'+id_{xy}$, and $s'+id_{x^2-y^2}$ pairings are nearly identical at $\Delta_0=0.26$meV, while for the $d+ig$ and $d_{x^2-y^2}$ pairings, $\Delta_0$ is found to be $0.14$meV.
%
%
%

%
%%%%%%%%%%%%%%%%%%%%====== figure ======%%%%%%%%%%
\begin{figure}
\hspace*{\fill}%
{%
  \includegraphics[width=1\columnwidth]{Tc}%
}\hspace*{\fill}%
\caption{The critical temperature versus (a) the chemical potential $\mu$ and (b) the $z$-direction hopping $g^\prime$ for the variety of pairings. The increase in $\mu$ grows up the filling of the band $\gamma$ leading to the charge transfer from the orbitals $d_{xz/yz}$ to the orbital $d_{xy}$ while the coupling $g^\prime$ decreases the filling of the band $\gamma$. The dash-dotted vertical line shows the critical chemical potential. These plots can be criterion for choosing the favored pairing.} \label{mug}
\end{figure}
%
%%%%%%%%%%%%%%%%%%%%%%%%%%%%%%%%%%%%%%%%%%%%%%%%%%
%

To facilitate comparison between different pairing states, we examine the effect of $g^\prime$ and $\mu$ on the critical temperature $T_c$.
This analysis helps narrow down possible pairing channels and identify the favored pairing for the d$_{xy}$ orbital. The results show a pronounced sensitivity of $T_c$ to increases in $g^\prime$ and $\mu$ across various pairings, as illustrated in Fig. \ref{mug}. For the $d+ig$ and $d_{x^2-y^2}$ pairings, $T_c$  reaches a significant peak, more than doubling, as it approaches the critical chemical potential $\mu_c=0.148$ meV. This critical point corresponds to a Lifshitz transition, characterized by a Fermi surface reconstruction where the vHS intersect the Fermi level at four points $(k_x,k_y)=(0,\pm\pi)$
and $(\pm\pi,0)$ within the Brillouin zone. Beyond $\mu_c$, $T_c$ declines rapidly as a function of the particle concentration in the d$_{xy}$ orbital.
In contrast, for the $p+ip$, $s^\prime+id_{xy}$, and $s^\prime+d_{x^2-y^2}$ pairings, $T_c$ exhibits little sensitivity to the vHS, with no peak
near $\mu_c$. This difference arises because the gap functions of $p+ip$ and $s^\prime+id_{xy}$ vanish at the vHS points $(k_x,k_y)=(0,\pm\pi)$
and $(\pm\pi,0)$, while the $s^\prime+d_{x^2-y^2}$ pairing is purely imaginary at these points. By comparison, the favored pairings, $d+ig$
and $d_{x^2-y^2}$, maintain real-valued gap functions.
%\\

As shown in Fig.~\ref{mug}(b), the influence of the coupling parameter $g^\prime$ on $T_c$
is consistent across the $d+ig$, $d_{x^2-y^2}$, and $p+ip$ pairings: increasing $g^\prime$
reduces $T_c$ by shifting the peak of the $d_{xy}$ orbital DOS away from the Fermi level.
This trend contrasts with the effect of $\mu$, which generally enhances $T_c$ by moving the
DOS peak closer to the Fermi level. For the $s^\prime+id_{xy}$ and $s^\prime+d_{x^2-y^2}$
pairings, however, $T_c$ shows anomalous behavior, beginning to increase once
$g^\prime \gtrsim 4$ meV.
Taken together with elastocaloric data indicating
that the superconducting gap remains finite at the van Hove points~\cite{li2022elastocaloric},
and considering that the effect of $g^\prime$ on the spectrum and therefore on $T_c$ is
opposite to that of $\mu$, we conclude that the most plausible pairing channels for the
$d_{xy}$ orbital are $d+ig$ and $d_{x^2-y^2}$.

%

%%%%%%%%%%%%%%%%%%%%%%%%%%%%%%%%%%%%%%%%%%%%%%%%%%%%%%%%%%%%%%%%%%%%%%%%%%%%%%%%%%%%%%%%%%%%%%%%%%%%%%%%%%%%%
%%%%%%%%%%%%%%%%%%%%%%%%%%%%%%%%%%%%%%%%%%%%%%%%%%%%%%%%%%%%%%%%%%%%%%%%%%%%%%%%%%%%%%%%%%%%%%%%%%%%%%%%%%%%%%%%%%%%%
%%%%%%%%%%%%%%%%%%%%%%%%%%%%%%%%%%%%%%%%%%%%%%%%%%%%%%%%%%%%%%%%%%%%%%%%%%%%%%%%%%%%%%%%%%%%%%%%%%%%%%%%%%%%%%%%%%%%
%
\section{Optical Hall conductivity}
\label{OPH}
%OPHPKE
Reformulating the superconducting Hamiltonian with a $6 \times 6$ matrix to account for the 2 Nambu and 3 orbital degrees of freedom, the bare Green's function in the superconducting state is given by
%
%
%========================================================
%========================Equation========================
%========================================================
\begin{align}
\label{G-func}
{\hat {\cal G}}_0(\bk,{ i  }{\upsilon} _n)
=\mqty*[\hat{G}_0(\bk,{ i  }{\upsilon} _n)
&
\hat{F}_0(\bk,{ i  }{\upsilon} _n)
\\
\hat{F}^\dag_0(\bk,{ i  }{\upsilon} _n)
&
-\hat{G}^{\intercal}_0(-\bk,-{ i  }{\upsilon} _n)],
\end{align}
%==============================================================
%
%
in which
$\hat{G}_0(\bk,{ i }\omega_m)$
and
$\hat{F}_0(\bk,{ i }\omega_m)$
are the normal and anomalous Green's functions, respectively.
%\\
%
%
Within the linear response theory,
the dynamical Hall conductivity is given by the following Kubo formula
%
%=============================================================
%========================Equation=============================
%=============================================================%
\be
\bl
\label{hh}
\sigma^H(\omega)&=\frac{1}{2}  \lim_{\bq \to 0}[\sigma^{}_{x,y}(\bq,\omega)-\sigma^{}_{y,x}(\bq,\omega)],
\el
\ee
where
\be\no
\bl
\sigma^{}_{i,j}(\bq,\omega)&=\frac{ i }{\omega}K^{}_{i,j}(\bq,\omega);
\quad
i,j\in \lbrace x,y \rbrace,
\el
\ee
%=============================================================
%
and $K^{}_{i,j}(\bq,\omega)$ is the  current-current correlation function that  is obtained from the analytical continuation,
${ i }\omega^{}_m\!\rightarrow\! \omega+{ i }0^+$, of its Matsubara counterpart
%
%=============================================================
%========================Equation=============================
%=============================================================
\be
\bl
\label{correlation}
&K^{}_{i,j}(\bq,{ i }\omega_m)
=
\\
&
%k_B
T \sum_{\bk,{ i } {\upsilon}_n}
{\Trace}
\Big[
{\hat {\cal J}}_i(\bk)
{\hat {\cal G}}_0(\bk,i {\upsilon} _n)
{\hat {\cal J}}_j(\bk)
{\hat {\cal G}}_0(\bk+\bq,
{ i }  {\upsilon}_n
\!+\!
{ i }{\omega}_m)
\Big].
\el
\ee
%=============================================================
Here, ${\hat{\cal J}}(\bk)$ is the charge current operator represented as a
$6\times 6$-rank tensor, and ${\upsilon}_n = (2n + 1) \pi T$ and ${\omega}_m = 2m\pi T$ are the Matsubara fermionic and bosonic frequencies, respectively.
After performing straightforwards calculations~\cite{Yazdani2024Polar}, the final expression for the dynamical Hall conductivity can be expressed as
\be
\bl
\label{hall:conduc}
&\sigma^H(\omega)
=
\frac{e^2}{\hbar}
\sum_{\bk }\sum_{\substack{\nu,\nu^\prime,\nu^{\prime\prime}\\(\nu\neq\nu^\prime\neq\nu^{\prime\prime})}}
{\cal T}_{\nu,\nu^\prime}^{\bk }
\Im[
\Delta_{\nu}^{\bk}\Delta^{\bk *}_{\nu^\prime}
]
\times
\\
&
\hspace{1.cm}
\big[
v_{x\nu\nu^\prime}^{\bk}
(v_{y\nu}^{\bk}-v_{y\nu^\prime}^{\bk})
-v_{y\nu\nu^\prime}^{\bk}
(v_{x\nu}^{\bk}-v_{x\nu^\prime}^{\bk})
\big]
\Upsilon_{\nu^{\prime\prime}}(\bk ,i\omega)
,
\el
\ee
where
\be\bl
\label{function}
&
\Upsilon_{\nu}(\bk ,i\omega)
=
\\
&\hspace{0.5cm}
\sum_{z}
\frac{(z-z^\prime)(z+z^\prime)^2
[z^2-E_{\nu}(\bk )^2]
[{z^\prime}^2-E_{\nu}(\bk )^2]
}
{\prod_{i=1,2,3}
[z^2-E_i(\bk )^2][{z^\prime}^2-E_i(\bk )^2]
},
\el
\ee
in which $z={ i }\upsilon_n$ and $z^\prime=z+{ i }\omega_m$.
Here,
$v_{l\nu\nu^\prime}^{\bk}=\pdv*{
{\cal T}_{\nu,\nu^\prime}^{\bk }
}{k_l}$
and
$ v_{l\nu}^{\bk}=\pdv*{\xi^{\bk}_{\nu}}{k_l}$ ($l=x,y$) are interorbital velocity and intraorbital velocity, respectively, and in fact are the elements of  the $3\times3$ bare current vertex in the orbital basis.
Furthermore, $E_{\nu}(\bk )=\sqrt{\xi^{\bk2} _{\nu}+\Delta^{\bk2}_{\nu}}$,
and the BCS quasiparticle spectra $E_i(\bk)$ are  obtained via the relation
$$
\det\hat{\cal G}^{-1}_0(\bk,z)
\propto
\prod_{i=1,2,3}[z^2 - E_i(\bk)^2].
$$
After carrying out the Matsubara summation in the expression of Eq. (\ref{function}), it becomes
\bea
\bl
\label{function1}
&
\Upsilon_{\nu}(\bk ,i\omega)=
\\
&
\frac{
E_{\nu}(\bk )^4-E_{\nu}(\bk )^2
[
E_1(\bk )^2+E_2(\bk )^2
]
+[E_1(\bk ) E_2(\bk )]^2}
{
E_1(\bk )E_2(\bk )
[E_2(\bk )^2-E_3(\bk )^2][E_3(\bk )^2-E_1(\bk )^2]
}
\times
\\
&
\Big[
\frac{
f(E_1(\bk ))-f(E_2(\bk ))}{
[E_1(\bk )-E_2(\bk )]
\Big(
[E_1(\bk )-E_2(\bk )]^2-(i\omega)^2
\Big)}
\\
&+
\frac{1-f(E_1(\bk ))-f(E_2(\bk ))}{
[E_1(\bk )+E_2(\bk )]
\Big(
[E_1(\bk )+E_2(\bk )]^2-(i\omega)^2
\Big)}
\Big]
\\
&
+
\big(
\mathrm{1\longrightarrow2\:and\:2\longrightarrow3\big)+\big(1\longrightarrow3\:and\:2\longrightarrow1}\big),
\no
\el
\\
\eea
where $f(E_\nu(\bk ))$ refers to the finite-temperature Fermi–Dirac function.

%
%%%%%%%%%%%%%%%%%%%%%%%%%%%%%%%%%%%%%%%%%%%%%%%%%%
%%%%%%%%%%%%%%%%%%%%====== figure ======%%%%%%%%%%%%%%%%%%%
\begin{figure}[t]
\centering
\includegraphics[width=1\linewidth]{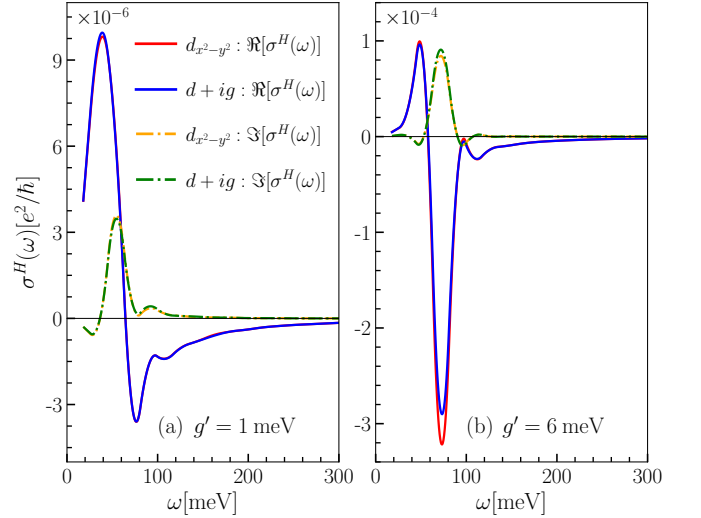}
\caption{%(Color online)
The comparison between the dynamical Hall conductivity,  $\sigma^H(\omega)$, for the pairing channels $d+ig$ and $d_{x^2-y^2}$ that indicates the role of the breaking of the time reversal symmetry on the gap function of the orbital $d_{xy}$ with
(a)  $g^\prime=1$meV, and
(b) $g^\prime=6$meV.
}
\label{conductivity}
\end{figure}
%%%%%%%%%%%%%%%%%%%%%%%%%%%%%%%%%%%%%%%%%%%%%%%%%%
%

Eq.~(\ref{hall:conduc}) states that to obtain a Hall-type response, at least one of the orbital
gap functions must be complex. The chiral character arises from the multiple order parameters
of the quasi-1D orbitals, and we set the gap functions $d+ig$ and $d_{x^2-y^2}$ on the $d_{xy}$ orbital.
Another requirement is the presence of a $\bk$-dependent inter-orbital coupling.
In Fig.~\ref{conductivity}, the dynamical Hall conductivity is shown for the favored pairings,
$d+ig$ and $d_{x^2-y^2}$, at two different coupling values, $g^\prime = 1$meV and
$g^\prime = 6$meV. From this, two main conclusions arise. First, the coupling parameter $g^\prime$
significantly influences the Hall conductivity. This effect is likely due to changes in the convex
and concave portions of the Fermi surfaces, which contribute positively and negatively to the
Hall conductivity, respectively~\cite{Ong1991Geometric}.
The induced concavity on the $\gamma$ band brings it closer to the $\beta$ band and
eventually into contact. This band proximity (nearly degenerate states) in the diagonal region,
which dominates the Berry curvature and Hall transport, causes both bands to contribute similarly.
As a result, the $\gamma$ band, which previously had negligible influence, now plays a prominent role.
This situation is analogous to the enhancement of electronic correlations in the quasi-1D
$d_{xz/yz}$ orbitals when they become nearly degenerate with the quasi-2D $d_{xy}$ orbital
\cite{Deng2016Transport}.
Secondly, the Hall response is found to be identical for both pairings, which indicates that
time-reversal symmetry breaking in the gap function of the $d_{xy}$ orbital does not affect the
Hall-type response in this context.
In fact, there is no need for a time-reversal-symmetry-breaking accidentally degenerate state, such as the $d + ig$ pairing. The dominant contribution instead arises from nearest-neighbor pairing in the $d_{xy}$ orbital, as suggested in Ref.~\cite{Yuan2023Multiband}.
The main contribution to the dynamical Hall conductivity comes from the term involving
${\cal T}_{xz,yz}^{\bk}$. Therefore, in the following we consider $d_{x^2-y^2}$ pairing for
the $d_{xy}$ orbital and focus on the term with inter-orbital coupling ${\cal T}_{xz,yz}^{\bk}$.
We also find that the dominant contribution to the Hall-type response originates from the
quasiparticle energy spectra $E_1(\bk)$ and $E_2(\bk)$, whereas the spectrum $E_3(\bk)$, which
is closest to the Fermi level, contributes only negligibly.

\begin{figure}[t]
\centering
\includegraphics[width= \linewidth]{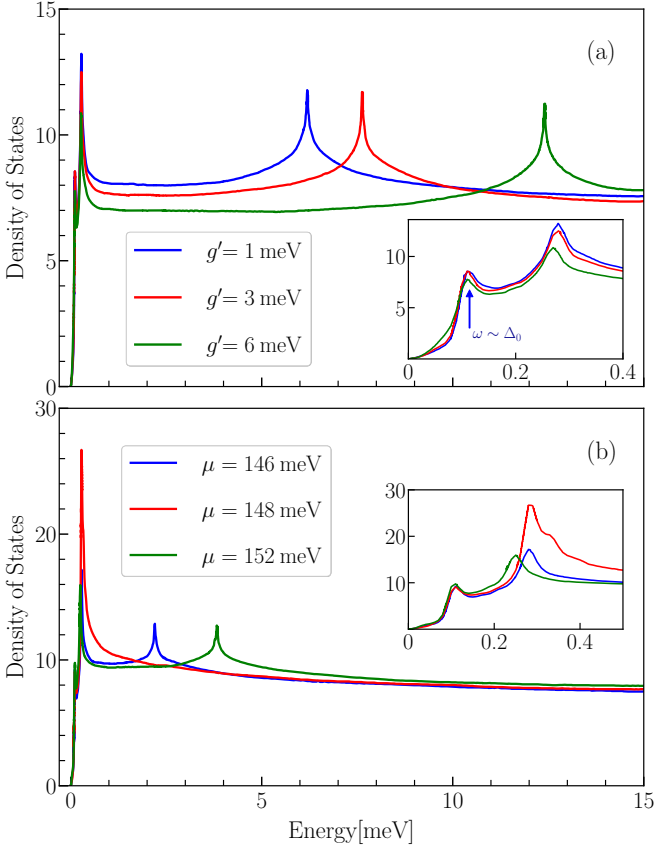}
\caption{
The DOS in the superconducting state is shown for the $p+ip$ wave pairing on the quasi-1D orbitals and the $d+ig$ wave pairing on the quasi-2D orbital at
$T=0$: (a) for different $g^\prime$ values with
$\mu = 142~\mathrm{meV}$, and (b) for different $\mu$ values with $g^\prime = 1~\mathrm{meV}$.
The insets show the same data with a zoomed-in view for the low-energy regimes.
}
\label{dos}
\end{figure}

A peak or resonance in the response function appears due to the coherence factor, which maintains a high density of states.
Therefore, considering peaks in both quantities is an essential step forward~\cite{Masuda2010Coherence}. Fig.~\ref{dos} shows the DOS in the superconducting state, with the $p+ip$ wave on the quasi-1D orbitals and the $d+ig$ pairing on the quasi-2D orbital, at zero temperature across different values of coupling $g^\prime$ and $\mu$.
At low energies, the DOS develops three main coherence peaks at $\omega \simeq \Delta_0$, $\omega \simeq 2\Delta_0$, and $\omega \simeq 8g^\prime$. The influence of interorbital coupling $g^\prime$ causes significant changes in the DOS: an increase in $g^\prime$ increases the intensity of the second peak and shifts the third peak to higher energies.
As the chemical potential $\mu$ increases, the first and second peaks are enhanced, and the third peak shifts to lower energies. At the critical chemical potential $\mu = \mu_c$, the third peak merges with the second peak. This critical point marks the crossing of van Hove singularities with the Fermi level, resulting in an increased DOS at low energies and the elimination of the second peak. With a further increase in $\mu$, the third peak reappears at lower energies, corresponding to the hole-like Fermi surface.
Notably, a fourth peak appears at higher energies, specifically at $\omega \simeq 4g$, which drives the primary resonance in the Hall-type response (see Fig.~\ref{conductivity}). This peak remains unaffected by variations in $g^\prime$ and $\mu$, and is therefore omitted from the present figure.

%
%%%%%%%%%%%%%%%%%%%%%%%%%%%%%%%%%%%%%%%%%%%%%%%%%%
%%%%%%%%%%%%%%%%%%%%====== figure ======%%%%%%%%%%%%%%%%%%%
%%%%%%%%%%%%%%%%%%%%%%%%%%%%%%%%%%%%%%%%%%%%%%%%%%
%%%%%%%%%%%%%%%%%%%%====== figure ======%%%%%%%%%%
%

The behavior of the coherence factor  in the $\sigma^H(\omega)$, i.e. the coefficient of function $\Upsilon_{\nu}(\bk ,i\omega)$ multiplied by function ${\cal T}_{\nu,\nu^\prime}^{\bk }
\Im[
\Delta_{\nu}^{\bk}\Delta^{\bk *}_{\nu^\prime}
]$~\cite{taylor2013anomalous}, is influenced by the proximity (touching) conditions between multiple Fermi surface sheets, which in turn control the peak positions in the response function.
This proximity is predominantly occurred in the diagonal regions ([110] and $[1\bar{1}0]$) rather than in the $[100]$ and $[010]$ directions, resulting in a significant reduction of the coherence factors along the $k_x$ and
$k_y$ axes.
In the absence of impurities
the multi-orbital nature of the system  primarily contributes to the observed Hall response.

%
%
%%%%%%%%%%%%%%%%%%%%%%%%%%%%%%%%%%%%%%%%%%%%%%%%%%
%%%%%%%%%%%%%%%%%%%%====== figure ======%%%%%%%%%%%%%%%%%%%
\begin{figure*}[!]
\centering
\includegraphics[width= 0.9\linewidth]{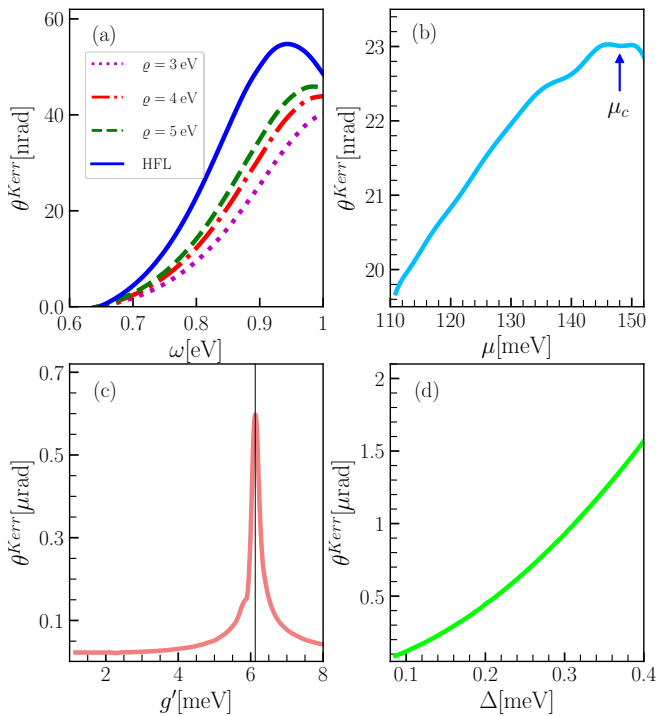}
\caption{%(Color online)
(a)
Kerr rotation angle (in unit of nanoradian) versus photon energy. The non-solid curves represent the influence of the interband electron-electron scattering on the Kerr angle in the hydrodynamic regime, $\varrho=1/\tau\gg1/\tau_\iota\sim1/\tau_\kappa$\textbf{}. The blue solid curve indicates the Kerr signal in the high frequencies limit (HFL), where the longitudinal conductivity is described by the Drude formula.
The Kerr polar angle is shown:
(b) as a function of the chemical potential $\mu$ with $g^\prime = 1$meV and $\Delta = 0.14$meV,
(c) as a function of the interorbital coupling or $z$-direction hopping $g^\prime$ with $\mu = 142$meV and $\Delta = 0.14$meV, we plotted it in the possible range from $0$ to $8$meV~\cite{Oda2019Paramagnetic}, and
(d) as a function of the gap function magnitude $\Delta$ with $g^\prime = 1$meV and $\mu = 142$meV.
We set $\omega =0.8$eV~\cite{Xia2006High}, although the general behavior remains consistent across other frequency ranges.
The maximum signal corresponds to the formation of near-Fermi-energy degenerate states between the $\beta$ and $\gamma$ bands. %
The region near the maximum in plot (c) corresponds to specific
$z$-direction hopping values that induce a concave curvature in the Fermi surface of the
 $\gamma$ sheet. 
The effect of SOC on the polar Kerr angle is shown for the upper pseudospin sector in panel (e), and for both pseudospin sectors (up and down) in panel (f).
 %Note that the plot corresponding to zero value of SOC in (e) is the plot (c).}{\color{red}
 The presence of SOC causes a decrease in the Kerr angle and the disappearance of the observed peak at higher SOCs}.

\label{ker}
\label{kerr}
\end{figure*}
%%%%%%%%%%%%%%%%%%%%%%%%%%%%%%%%%%%%%%%%%%%%%%%%%%
%%%%%%%%%%%%%%%%%%%%====== figure ======%%%%%%%%%%
%

%%%%%%%%%%%%%%%%%%%%%%%%%%%%%%%%%%%%%%%%%%%%%%%%%%%
%%%%%%%%%%%%%%%%%%%%%%%%%%%%%%%%%%%%%%%%%%%%%%%%%%
%%%%%%%%%%%%%%%%%%%%%%%%%%%%%%%%%%%%%%%%%%%%%%%%%%
\section{Polar Kerr Effect}
\label{PKE}
Using the standard formalism for Kerr rotation and incorporating the dynamical Hall conductivity, the polar Kerr angle can be evaluated as~\cite{Taylor2012Intrinsic}
%==========================================================
%=======================Equation===========================
%==========================================================
%
\be
\bl
\theta^{Kerr}(\omega)=
\frac{4\pi }{\omega d}
\Im
\Big[
\sigma^{H}(\omega)\varphi(\omega)
\Big],
\el
\ee
%==========================================================
%
 %
where
%==========================================================
%=======================Equation===========================
%==========================================================
%
$
1/\varphi(\omega)=
%\Big[
n(\omega)
[n(\omega)^2-1]
%\Big]
,
$
 and
$n(\omega)=\sqrt{\epsilon_\infty+(4\pi i/\omega)\sigma(\omega)}$ denotes the refraction index.
Here  $\epsilon_\infty=10$ refers to the background dielectric tensor, and $d=6.8${\AA} is the interlayer distance.
Obviously, the refraction index is defined by the complex longitudinal
optical conductivity $\sigma(\omega)$.
For simplicity,  without any loss of generality, the calculation can be performed using a two-orbital model~\cite{dora2003optical}.
In which, the first orbital represents the sum of quasi-1D orbitals, denoted as $d_{xz} + d_{yz}$, while the second orbital corresponds to the quasi-2D orbital  $d_{xy}$.
This model arises due to the tetragonal symmetry of Sr$_2$RuO$_4$ and the distinct orbital characteristics of the three Fermi surfaces. These features suggest that the dominant pairing channel originates from either the quasi-2D orbital $ d_{xy}$ or the quasi-1D orbitals  $d_{xz/yz}$~\cite{Raghu2013Theory}.
Following Refs.~\cite{maslov2016optical}, the conductivity is obtained by solving two coupled equations of motion for free charges, characterized by the velocities $\mathbf{V}\kappa$ and $\mathbf{V}\iota$ in the orbitals $\kappa$ and $\iota$, respectively. This leads to the expression
%==========================================================
%=======================Equation===========================
%==========================================================
%
\be
\bl
\label{motion}
-i\omega m_\iota \mathbf{V}_{\iota} &= -\frac{m_\iota \mathbf{V}_{\iota}}{\tau_{\iota}} + q_\iota \mathbf{E} - \mathfrak{g}_{ee} \, n_\kappa (\mathbf{V}_\iota - \mathbf{V}_\kappa),
\\
-i\omega m_\kappa \mathbf{V}_{\kappa} &= -\frac{m_\kappa \mathbf{V}_{\kappa}}{\tau_{\kappa}} + q_\kappa \mathbf{E} - \mathfrak{g}_{ee} \, n_\iota (\mathbf{V}_\kappa - \mathbf{V}_\iota),
\el
\ee
%==========================================================
where
$m_{\iota(\kappa)}$, $q_{\iota(\kappa)}$, $n_{\iota(\kappa)}$, and $\tau_{\iota(\kappa)}$ denote the effective mass, charge, particle density, and scattering time of the charge carriers in the orbital $\iota(\kappa)$, respectively. Moreover, $\mathbf{E}$ denotes the electric field, and $\mathfrak{g}_{ee}$ represents the interorbital electron-electron scattering, which becomes relevant only at non-zero temperatures.
We continue by calculating the velocities $\mathbf{V}_\iota(\kappa )$, which is given by
\be
\bl
&\mathbf{V}_{\iota(\kappa)}=
\frac{m_{\kappa(\iota)}q_{\iota(\kappa)}(\tau_{\kappa(\iota)}^{-1}-i\omega)+
\mathfrak{g}_{ee}
(n_\iota q_\iota+n_\kappa q_\kappa)}{m_\iota m_\kappa
\Big((\gamma_0^2-\omega^2)-\frac{i\omega}{\tau^\prime}\Big)}\mathbf{E},
\el
\ee
with
\be
\bl
&\frac{1}{\tau^\prime}=\frac{1}{\tau_\iota}+\frac{1}{\tau_\kappa}+\frac{1}{\tau},
\\
&\gamma_0^2=\frac{1}{\tau_\iota\tau_\kappa}+\frac{1}{\tau}\Big[\frac{1}{\tau_\iota(1+\frac{n_\kappa m_\kappa}{n_\iota m_\iota})}+\frac{1}{\tau_\kappa(1+\frac{n_\iota m_\iota}{n_\kappa m_\kappa})}\Big],
\\
&\frac{1}{\tau}=\mathfrak{g}_{ee}
\Big(\frac{n_\iota}{m_\kappa}+\frac{n_\kappa}{m_\iota}
\Big),
\el
\ee
After performing some calculations and comparing the relations for the current $\mathbf{J} = n_\iota q_\iota \mathbf{V}_\iota + n_\kappa q_\kappa \mathbf{V}_\kappa$ and $\mathbf{J} = \sigma(\omega) \mathbf{E}$, we obtain the conductivity $\sigma(\omega)$ as
\be
\bl
\label{chubukov}
\sigma(\omega)=\frac{1}{4\pi}\frac{\frac{\omega_{\iota p}^2}{\tau_\kappa}+\frac{\omega_{\kappa p}^2}{\tau_\iota}+\frac{\omega_0^2}{\tau}-i\omega\omega_p^2}{(\gamma_0^2-\omega^2)-\frac{i\omega}{\tau^\prime}},
\el
\ee
%==========================================================
%
where $\omega^2_{\iota p}=(n_\iota q^2_\iota)/m_\iota$ is the plasma frequency of the orbital $\iota$, $\omega^2_p=\omega^2_{\iota p}+\omega^2_{\kappa p}$
and
$\omega^2_0=4\pi(n_\iota q_\iota+n_\kappa q_\kappa)^2/(n_\iota m_\iota+n_\kappa m_\kappa)$.
Note that $\tau$ only depends on the temperature not frequency.
Considering  a generic case, when $n_{\iota}\sim n_{\kappa}$ and $m_{\iota}\sim m_{\kappa}$ which implies that $\omega_{\iota p}\sim\omega_{\kappa p}\sim\omega_p\sim\omega_0=2.9$eV and $\tau^{-1}_\iota\sim\tau^{-1}_\kappa =0.4$eV~\cite{maslov2016optical},
one can focus on two regimes:
$\mathrm{(i)}$ the high-frequency limit, where the longitudinal conductivity is described by the Drude formula, and
$\mathrm{(ii)}$ the hydrodynamic regime, characterized by the condition $\varrho = \tau^{-1} \gg \tau^{-1}_\iota \sim \tau^{-1}_\kappa$~\cite{maslov2016optical}.
In this context, Fig.~\ref{kerr}(a) presents the Kerr rotation angle (in units of nanoradians) as a function of photon energy.
The blue-solid curve represents the Kerr signal in the high-frequency limit.
The remaining curves correspond to $e$-$e$ scattering for $\varrho = 3$, $4$, and $5$eV, which are relevant in the hydrodynamic regime and highlight the effect of interband electron-electron scattering on the Kerr angle.
We observe that while scattering enhances the Kerr polar angle, the enhancement remains smaller than that observed in the high-frequency limit.
Moreover, we  plot  the Kerr polar angle as a function of the chemical potential in the Fig.~\ref{kerr}(b), the $z$-direction hopping in the Fig.~\ref{kerr}(c), and the pairing amplitude in the Fig.~\ref{kerr}(d), all at the frequency  $\omega= 0.8eV$~\cite{Taylor2012Intrinsic}.
Fig.~\ref{kerr}(b) demonstrates that as the chemical potential increases, $\theta^{Kerr}(\omega)$ rises until it crosses the van Hove point $\mu = \mu_c$, where a Lifshitz transition occurs. After this point, the superconducting character of the $\gamma$ band becomes fragile.
As seen in Fig.~\ref{kerr}(c), increasing the coupling $g^\prime$ leads to an increase in the Kerr angle. This is expected, as increasing $g^\prime$ is equivalent to transferring an electron from the 2D quasi-orbital to the 1D quasi-orbitals, and it is anticipated that this transfer will result in an increase in the Kerr angle.
 A sharp peak in the Kerr angle is observed at  $g^\prime\simeq6$meV,
where the touching between the bands $\beta$ and $\gamma$ is largest and the nearly degenerate low-energy electronic states are formed, i.e., the change in band dispersion in vicinity of the Fermi energy. In fact, the region around the peak, % shown by the cyan colored area,
is related to the appearance of the concavity on the band $\gamma$.
Fig.~\ref{kerr}(d) illustrates the dependence of the Kerr angle on the pairing amplitude, highlighting the role of temperature in the Kerr angle. As temperature increases, the amplitude of the gap function decreases, which in turn reduces the Kerr angle.

To complete the analysis, we now examine the role of SOC in the
polar Kerr effect. Figure~\ref{kerr}(e) shows the SOC dependence of
the Kerr response for the pseudospin model, $\theta^{\mathrm{Kerr}}$. SOC reduces
the overall signal and suppresses the pronounced peak observed in
Fig.~\ref{kerr}(c). The net outcome is that SOC lifts the near-touching
between the $\beta$ and $\gamma$ bands, thereby weakening the formation of
nearly degenerate states and diminishing the Kerr response. Nevertheless,
the underlying mechanism persists: band proximity still enhances the Hall
and Kerr responses, albeit with much smaller magnitude when SOC is present.
A similar behavior is found when SOC is incorporated in the {\it full
multi-orbital BdG (Nambu) framework}, i.e., by combining the contributions
from both pseudospin sectors, see Fig.~\ref{kerr}(f).
Including SOC removes the contributions to $\sigma^{H}(\omega)$ associated
with the coefficients $\Im\!\left[\Delta_{\nu}^{\bk}\Delta_{xy}^{\bk *}\right]$
$(\nu=xz,yz)$, while
simultaneously enhancing the term proportional to
$\Im\!\left[\Delta_{xz}^{\bk}\Delta_{yz}^{\bk *}\right]$. These arise from the singlet pairing on the $d_{xy}$ orbital
coupled with triplet pairing on the $d_{xz}$ and $d_{yz}$ orbitals.

At the end, we should emphasize that the real part of the dynamical Hall conductivity follows a similar trend to the Kerr polar angle. This similarity is based on the fact that the imaginary part of the Hall-type response vanishes
 at high frequencies, as shown in the  Fig.~\ref{conductivity}.
%
%

%%%%%%%%%%%%%%%%%%%%%%%%%%%%%%%%%%%%%%
%%%%%%%%%%%%%%%%%%%%%%%%%%%%%%%%%%%%%%
\section{Conclusion}
We have investigated the Hall-type response of the superconductor Sr$_2$RuO$_4$ in two physical scenarios. The first scenario involves inducing a Lifshitz transition in the largest Fermi surface sheet, causing it to reshape from an electron-like to a hole-like (open sheet) configuration as the chemical potential, $\mu$, is increased.
The second scenario explores the impact of retracting the previously mentioned sheet from the van Hove points by adjusting the coupling between the quasi-1D and quasi-2D orbitals, i.e., the $z$-direction hopping parameter $g^\prime$.
Initially, we found $d+ig$ and $d_{x^2-y^2}$ pairing channels as the leading candidates for the $d_{xy}$ orbital in the weak-coupling limit. These two channels exhibit the correct and expected behaviors as $\mu$ and $g^\prime$ vary such that tuning the chemical potential $\mu$ leads to increasing the superconducting transition temperature whereas $g^\prime$ results in a reverse effect.
Using a self-consistent approach, we have determined the magnitude of these pairings at the zero-temperature limit around $\Delta_0 = 0.14$meV.
Furthermore, we have discovered that the optical response for these two pairings are identical. This finding suggests that the time-reversal symmetry breaking of the gap function in the $d_{xy}$ orbital is not required to produce a finite Kerr angle.

Similar to the two-orbital model, the primary peak in the response function arises in vicinity of the coupling amplitude between the quasi-1D orbitals.
Increasing $g^\prime$ affects the frequency region around this peak, leading to negative conductivity, similar to the effect of increasing $\mu$.
These effects are examined using the coherence factor. We have found that $g^\prime$ shifts the superconducting DOS peak to higher energies, while $\mu$ pushes it to lower energy levels.
Finally, we have studied the effects of electron-electron scattering, the gap function amplitude (or equivalently temperature), $g^\prime$, and $\mu$ on
the Kerr angle.
The effect of the former case is an enhancement of the Kerr signal in the high-frequency limit. As expected, the Kerr angle increases with the gap amplitude.
Additionally, the key effect of $g^\prime$ occurs around $\sim 6$meV, where a sharp peak emerges, corresponding to the increase in the orbital weight of quasi-1D orbitals. This is accompanied by the formation of near-Fermi energy degenerate states between the $\beta$ and $\gamma$ bands, resulting from the enhanced dispersive state of the $\gamma$ band near the Fermi level.
The presence of the spin-orbit coupling suppresses such degenerate states and gives rise to a reduction in the polar Kerr angle.
Moreover, the chemical potential, $\mu$, enhances the Kerr angle until it reached its critical value.
In conclusion, the Hall response can be tuned by adjusting various parameters, such as transferring electrons to the quasi-1D orbitals and enhancing superconductivity with increasing chemical potential, provided that the system does not undergo a Lifshitz transition that leads to an open geometry.
Additionally, these two parameters significantly influence the superconducting transition temperature by modifying the proximity of the Fermi surfaces associated with the $\beta$ and $\gamma$ bands.

\bibliography{References}
%%%%%%%%%%%%%%%%%%%%%%%%%%%%%%%%%%%%%%
%%%%%%%%%%%%%%%%%%%%%%%%%%%%%%%%%%%%%%%%%%%%%%%%%%%%%%%%%%%
\end{document}